\documentclass[twocolumn,aps,superscriptaddress,prl,floatfix,showpacs]{revtex4}
\usepackage{amsmath,amssymb}
\usepackage{graphicx}
\begin{document}
\title{Percolation with Multiple Giant Clusters}
\author{E.~Ben-Naim}\email{ebn@lanl.gov}
\affiliation{Theoretical Division and Center for Nonlinear
Studies, Los Alamos National Laboratory, Los Alamos, New Mexico
87545}
\author{P.~L.~Krapivsky}\email{paulk@bu.edu}
\affiliation{Center for Polymer Studies and Department of Physics,
Boston University, Boston, Massachusetts 02215}
\begin{abstract}
We study the evolution of percolation with freezing. Specifically, we
consider cluster formation via two competing processes: irreversible
aggregation and freezing. We find that when the freezing rate exceeds
a certain threshold, the percolation transition is suppressed. Below
this threshold, the system undergoes a series of percolation
transitions with multiple giant clusters (``gels'') formed. Giant
clusters are not self-averaging as their total number and their sizes
fluctuate from realization to realization. The size distribution
$F_k$, of frozen clusters of size $k$, has a universal tail, $F_k\sim
k^{-3}$. We propose freezing as a practical mechanism for controlling
the gel size.
\end{abstract}

\pacs{82.70.Gg, 02.50.Ey, 05.40.-a}

\maketitle

Percolation was originally discovered in the context of polymerization
and gelation \cite{flory,stock}. Percolation has found numerous
applications in physics \cite{ah,bh1}, geophysics \cite{sah},
chemistry \cite{floryB}, and biology \cite{pg}. It plays an important
role in a vast array of natural and artificial processes ranging from
flow in porous media \cite{BH} and cloud formation \cite{drake,sein}
to evolution of random graphs \cite{jlr,gg}, combinatorial optimization
\cite{hart}, algorithmic complexity \cite{bbckw}, amorphous computing,
and DNA computing using self-assembly \cite{win}.

We study the evolution of percolation using the framework of
aggregation.  An aggregation process typically begins with a huge
number of molecular units (``monomers'') that join irreversibly to
form clusters (``polymers'').  At some time, a giant cluster (``gel'')
containing a finite fraction of the monomers in the system is born,
and it grows to engulf the entire system. In this classic percolation
picture only a single gel forms, but in many natural and artificial
processes the system freezes into a non-trivial final state with
multiple gels or even micro-gels \cite{ms,sv}.  In this Letter, we
show that aggregation with freezing naturally lead to formation of
multiple gels and that freezing is also a convenient mechanism for
controlling the gel size.

We analyze the simplest aggregation with freezing process where there
are two types of clusters: active and frozen. Active clusters join by
binary aggregation into larger active clusters.  The aggregation rate
is proportional to the product of the two cluster sizes
\cite{drake,fran,aal}; this is equivalent to the gelation model of
Flory and Stockmayer where a chemical bond between two monomers joins
their respective polymers \cite{flory,stock}. In parallel, active
clusters may become frozen at a size-independent constant rate
$\alpha$.  These frozen clusters are passive, that is they do not
interact with other (active or passive) clusters.

This process is conveniently studied using the rate equation
approach. The density $c_k(t)$ of active clusters of mass $k$ at time
$t$ (that is, made up from $k$ monomers) satisfies
\begin{equation}
\label{ck-eq}
\frac{d c_k}{dt}=\frac{1}{2}\sum_{i+j=k} (ic_i)(jc_j) -m\,kc_k-\alpha\,c_k, 
\end{equation}
with $m(t)$ the mass density of active clusters.  The first two terms
on the right-hand side describe how the cluster size distribution
changes due to aggregation, and the last term accounts for loss due to
freezing. The quantity $m(t)$ is subtle.  Generally, it equals the
mass density of {\em all} active clusters including possibly giant
clusters, but when there are no giant clusters, i.e., all clusters are
finite in size, then $m(t)\equiv M_1(t)$ with the moments defined via
$M_n(t)\equiv \langle k^n\rangle =\sum_{k\geq 1} k^nc_k(t)$. We are
interested in the evolution starting with finite clusters only.

\smallskip
\noindent{\it The gelation transition.} Initially, all clusters are
finite in size, so $m=M_1$.  The moments $M_n$ provide a useful probe
of the dynamics.  From the governing equation (\ref{ck-eq}), the
second moment of the size distribution $M_2$ obeys the closed equation
$dM_2/dt=M_2(M_2-\alpha)$. For arbitrary initial condition, 
\begin{equation}
\label{M2-sol}
M_2(t)=\alpha\,\left[\left(\frac{\alpha}{\alpha_c}-1\right)\,
e^{\alpha t}+1\right]^{-1}.
\end{equation}
There is a critical freezing rate $\alpha_c=M_2(0)$.  For fast
freezing, $\alpha\geq \alpha_c$, the second moment is always finite
indicating that clusters remain finite at all times. In this case,
there is no gelation. For slow freezing, $\alpha<\alpha_c$, there is a
finite time singularity indicating that an infinite cluster, the gel,
emerges in a finite time \cite{finite}. The gelation time is
\begin{equation}
\label{tg} t_g=-\frac{1}{\alpha}\ln \left(1-\frac{\alpha}{\alpha_c}\right).
\end{equation}
The gelation point marks two phases. Prior to the gelation point, the
system contains only finite clusters that undergo cluster-cluster
aggregation (``coagulation phase''). Past the gelation point, the gel
grows via cluster-gel aggregation (``gelation phase''). We analyze
these two phases in order.

\smallskip
\noindent{\it Coagulation phase.}  Coagulation occurs for
$\alpha\geq\alpha_c$ at all times or for $\alpha<\alpha_c$ when
$t<t_g$. From (\ref{ck-eq}), the mass density of active clusters
satisfies $dm/dt=-\alpha\, m$, and thus, ordinary exponential decay
occurs,
\begin{equation}
\label{M1-sol}
m(t)=m(0)e^{-\alpha t}\,.
\end{equation}
For concreteness, we consider the monodisperse initial conditions
$c_k(0)=\delta_{k,1}$. In this case $M_n(0)=1$ and consequently,
$\alpha_c=1$. The cluster size distribution is obtained using the
transformed distribution, $c_k=e^{-\alpha t}\,C_k$, and the modified
time variable
\begin{equation}
\label{tau}
\tau=\int_0^t dt' e^{-\alpha t'}=\frac{1-e^{-\alpha t}}{\alpha}
\end{equation}
that increases monotonically with the physical time and reaches
$\tau\to 1/\alpha$ as $t\to\infty$. With these transformations,
Eq.~(\ref{ck-eq}) reduces to the no-freezing case
\begin{equation}
\label{ck-eq1}
\frac{d C_k}{d\tau}=\frac{1}{2}\sum_{i+j=k} (iC_i)(jC_j) - kC_k.
\end{equation}
From the well-known solution of this equation \cite{ziff,bk}, the
cluster-size distribution is
\begin{equation}
\label{ck-pre}
c_k(t)=\frac{k^{k-2}}{k!}\,\,\tau^{k-1}\,e^{-k\tau-\alpha t}.
\end{equation}
Generally, the size distribution decays exponentially at large sizes
and the typical cluster size is finite. The gelation time (\ref{tg})
is simply $\tau_g=1$. Of course, no gelation occurs when $\alpha>1$
because $\tau<1/\alpha<1$. Otherwise, as the gelation point is
approached, $t\to t_g$, the characteristic cluster size diverges
$k\sim (t_g-t)^{-2}$. The gelation point is marked by an algebraic
divergence of the size distribution $c_k\sim (1-\alpha)k^{-5/2}$ for
large $k$. We note that the mass density decreases linearly with the
modified time, $m=1-\alpha \tau$, and that at the gelation point, the
mass is simply $m(\tau_g)=1-\alpha$.

\smallskip
\noindent{\it Gelation phase.} Past the gelation transition, a giant
cluster containing a finite fraction of the mass in the system
forms. In addition to cluster-cluster aggregation, cluster-gel
aggregation takes place with the giant cluster growing at the expense
of finite clusters. In parallel, all clusters may undergo freezing and
particularly, the gel itself may freeze.

Formally, the size distribution (\ref{ck-pre}) generalizes to
\begin{equation}
\label{ck-post}
c_k(t)=\frac{k^{k-2}}{k!}\,\,\tau^{k-1}\,e^{-ku -\alpha t}
\end{equation}
with $u(t)=\int_0^t dt'\,m(t')$. Statistical properties of the size
distribution are derived from the generating function
$c(z,t)=\sum_{k\geq 1} kc_k(t)e^{kz}$ that equals
\begin{equation}
\label{czt}
c(z,t)=\tau^{-1}e^{-\alpha t} G(z+\ln\tau -u)
\end{equation}
where $G(z)=\sum_{k\geq 1}\frac{k^{k-1}}{k!}\,\,e^{kz}$ is the ``tree''
function \cite{jklp}. 

\begin{figure}[t]
 \includegraphics*[width=0.45\textwidth]{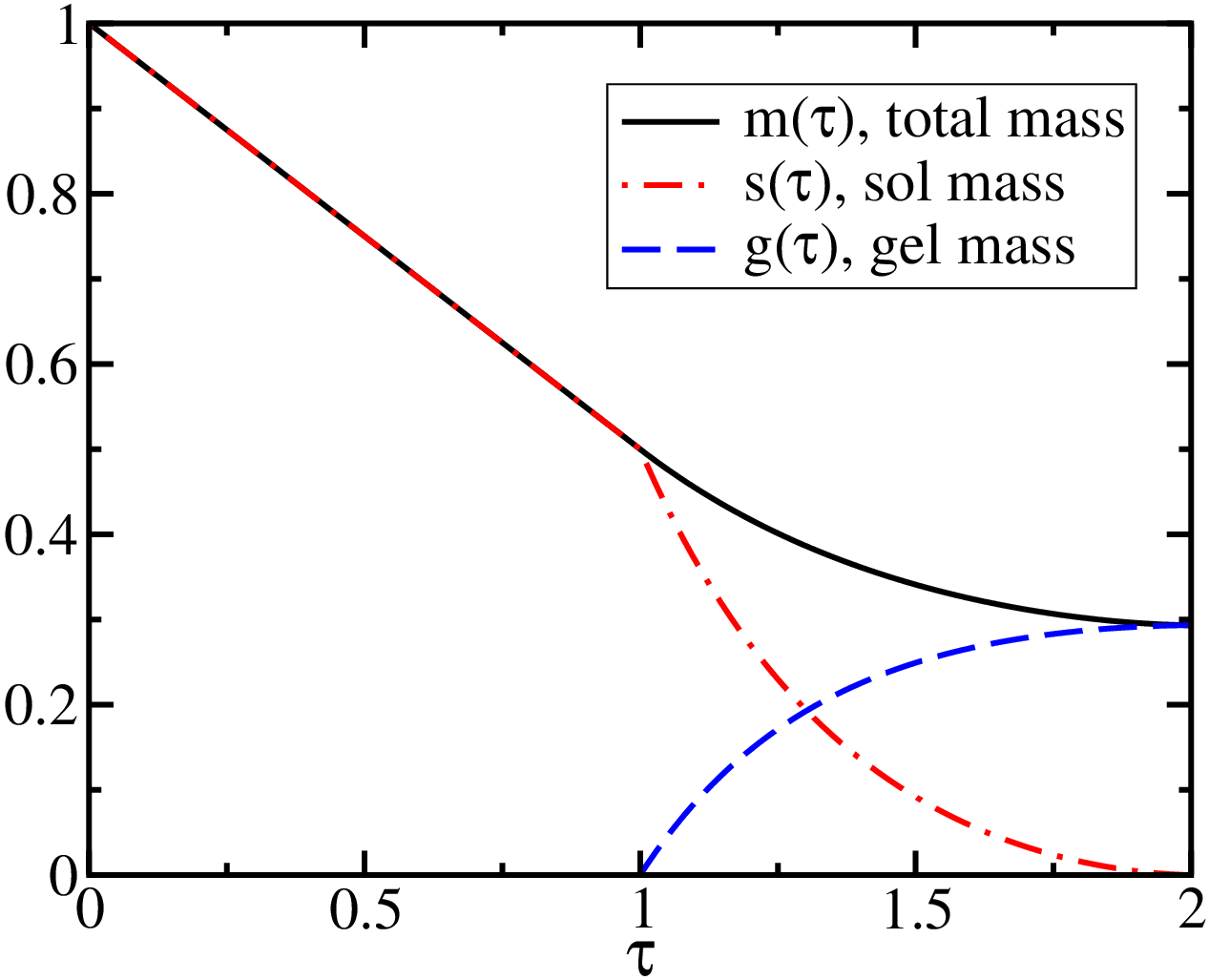}
 \caption{The total mass, the sol mass, and the gel mass versus
 modified time $\tau$ for $\alpha=1/2$. These curves hold as long as
 the gel remains active.}
\label{sol} 
\end{figure}

During the gelation phase, active clusters consist of finite clusters,
the ``sol'', with mass $s$, and the gel with mass $g$.  The total mass
density of clusters, $m=s+g$, decays according to
\begin{equation}
\label{mt-eq}
\frac{d m}{dt}= - \alpha s.
\end{equation}
The sol mass decays according to $ds/dt=-g\,M_2-\alpha s$ obtained
from (\ref{ck-eq}) using $s=M_1$. The first two moments follow from
the generating function, $M_1=c(0)$ and $M_2=c'(0)$, and using the
identity \hbox{$G'(z)=G/(1-G)$} \cite{wilf}, yields
\hbox{$M_2=s/(1-s\tau e^{\alpha t})$}.  The evolution equation for the
sol mass becomes explicit, 
\begin{equation}
\label{st-eq}
\frac{ds}{dt}=-\frac{s(m-s)}{1-s\tau e^{\alpha t}}-\alpha\,s\,.
\end{equation}
Equations (\ref{mt-eq}) and (\ref{st-eq}) are subject to the initial
conditions $m(t_g)=s(t_g)=1-\alpha$. Once the masses are found, the
formal solution (\ref{ck-post}) becomes explicit.  Results of
numerical integration of Eqs.~(\ref{mt-eq}) and (\ref{st-eq}) are
shown on Fig.~\ref{sol}. In the vicinity of the gelation transition,
the gel mass grows linearly with time:
\begin{equation}
\label{near} g(t)\simeq 2(1-\alpha)^2\,(t-t_g)
\end{equation}
as $t\downarrow t_g$. The quadratic dependence on the freezing rate
implies that the emerging gel is very small when $\alpha\uparrow
\alpha_c$. Thus, micro-gels, that may be practically indistinguishable
from large clusters, emerge. Moreover, the maximal gel size must be
smaller than $1-\alpha$. We conclude that freezing can be used to
control the gel size, as gels of arbitrarily small size can be
produced using freezing rates just below criticality.

\begin{figure}[t]
\includegraphics*[width=0.47\textwidth]{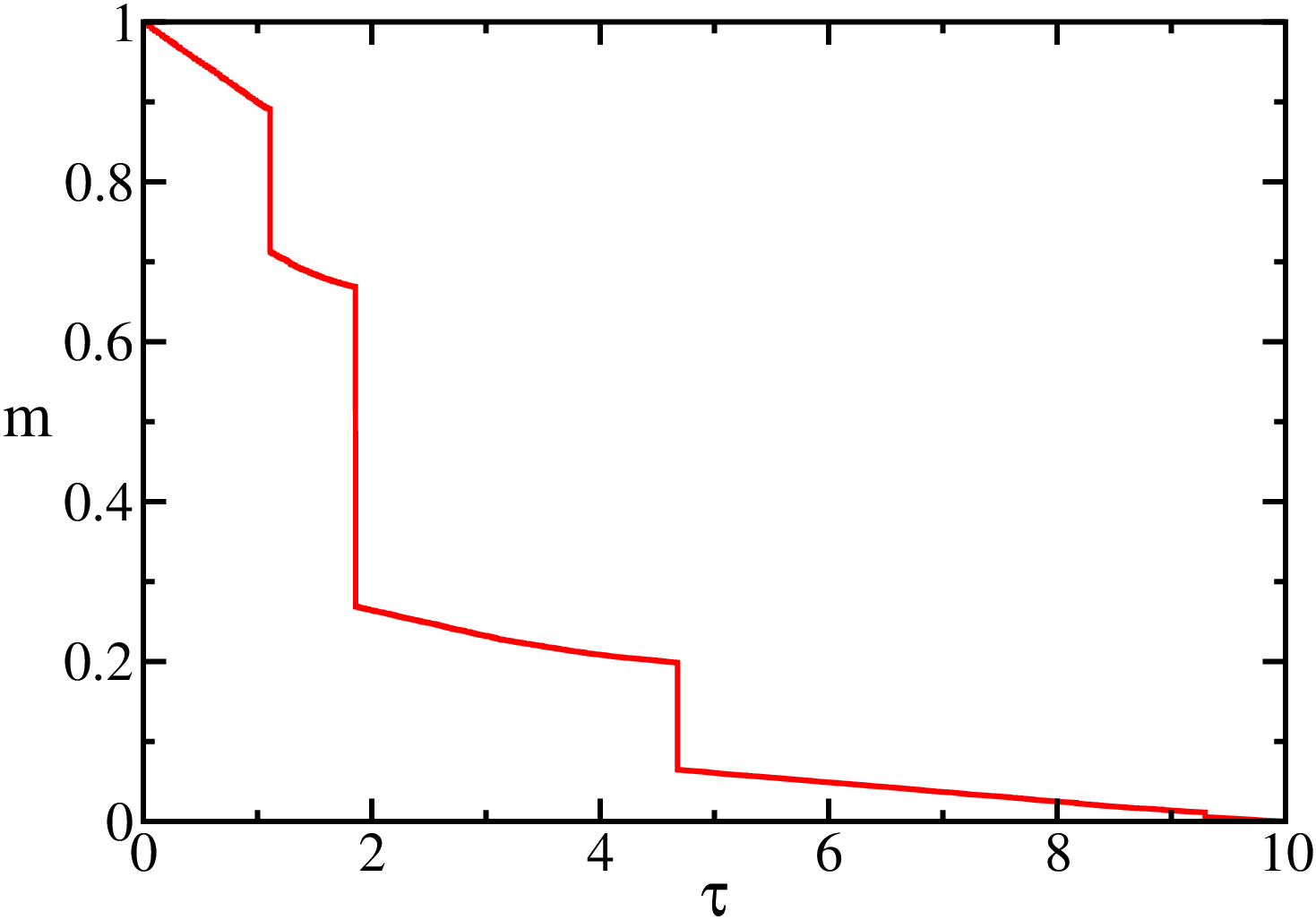}
\caption{The mass density $m$ versus time $\tau$. Shown are results of
a Monte Carlo simulation with system size $N=10^5$ and freezing rate
$\alpha=0.1$. The system alternates between the coagulation phase and
the gelation phase. In the former phase the mass decreases linearly
according to (\ref{M1-sol}) such that depletion occurs at time
$\tau=1/\alpha$. In the latter phase, the active mass decreases slower
than linear according to (\ref{mt-eq}) and (\ref{st-eq}).  The
gelation phase ends when the gel freezes.}
\label{mass}
\end{figure}

\smallskip
\noindent{\it Multiple giant clusters.} At any time during the
gelation phase, the gel itself may freeze. This freezing process is
random: the gel lifetime $T$ is a random variable that is
exponentially distributed, $P(T)=\alpha e^{-\alpha T}$. Until the gel
freezes, the system evolves deterministically, so the mass of the
frozen gel is $g(t_g+T)$. When the gel freezes, the total active mass
$m(t)$ is discontinuous: it exhibits a downward jump
(Fig.~\ref{mass}).  Given that the duration of the gelation phase is
governed by a random process, the mass of the frozen gel is also
random. It fluctuates from realization to realization, i.e., it is not
a self-averaging quantity.

When the gel freezes, the system re-enters the coagulation phase
because all remaining clusters are finite. The initial conditions are
dictated by the duration of the preceding gelation phase and are
therefore also stochastic. Nevertheless, once the initial state is
set, the evolution in the coagulation phase is deterministic. The gel
is frozen so it no longer affects the evolution and only
cluster-cluster aggregation occurs. Let us assume that the gel freezes
at time $t_f$. Reseting time to zero, the first and the second moments
are simply given by Eqs.~(\ref{M1-sol}) and (\ref{M2-sol}),
respectively, with $M_n(0)$ replaced by $M_n(t_f)$. Note that
$M_n(t_f)$ contains contributions from {\em finite} clusters only. A
second gelation occurs if the freezing rate is sufficiently small,
$\alpha<M_2(t_f)$. Otherwise, the system forever remains in the
coagulation phase.

\smallskip
\noindent{\it Cyclic dynamics.} The general picture is now clear: the
process starts and ends in coagulation and throughout the evolution,
the system alternates between coagulation and gelation.  Once the
initial conditions are set, the behavior throughout the coagulation
phase and throughout the gelation phase are both deterministic.  Each
gelation phase ends with freezing of the active gel. Since the
duration of the gelation phase is random, the size of the giant
clusters, and their number are both random variables.  Generically,
the system exhibits a series of percolation transitions, each
producing a frozen gel, so that overall, multiple gels are produced.
The random freezing process governs the number of percolation
transitions as well as the size of the frozen gels.

The magnitude of the second moment when the gel freezes determines
whether a successive gelation occurs. Because the second moment
diverges at the gelation point, there is a time window past the
gelation time where the second moment exceeds the freezing rate. If
the gel freezes during this window, another percolation transition is
bound to occur. Therefore, the maximal number of frozen gels is
unbounded.

Monte Carlo simulations confirm this picture (Fig.~\ref{mass}). In the
simulations, we keep track the total aggregation rate
$R_a=N(M_1^2-M_2)/2$ and the total freezing rate $R_f=\alpha
NM_0$. Aggregation occurs with probability $R_a/(R_a+R_f)$, and
freezing occurs with the complementary probability. A cluster is
chosen for aggregation with probability proportional to its size.
Time is augmented by $\Delta t=1/(R_a+R_f)$ after each aggregation or
freezing event.

\begin{figure}[t]
 \includegraphics*[width=0.46\textwidth]{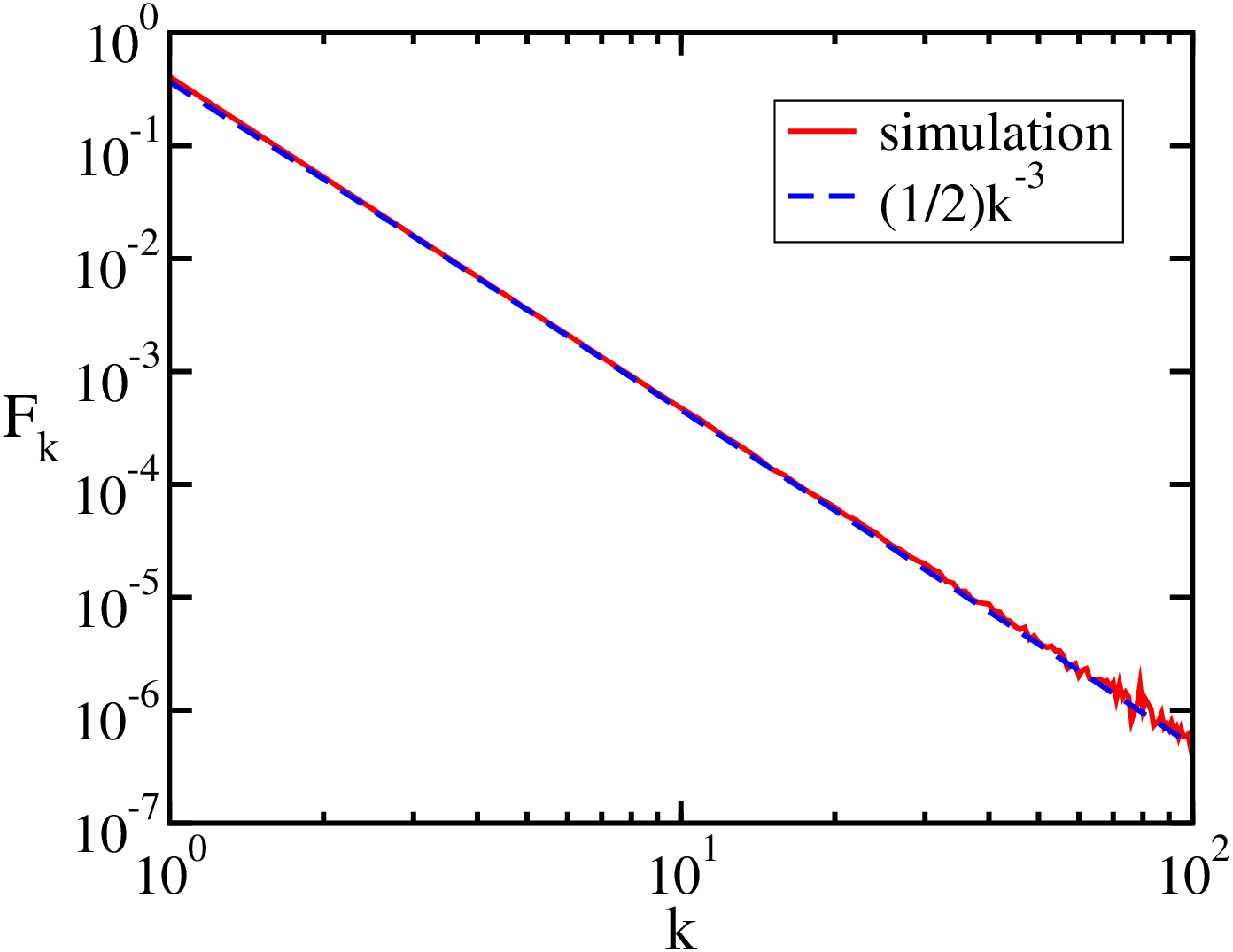}
 \caption{The size distribution of the frozen clusters for
   $\alpha=1/2$. The simulation results represent an average over
   $10^2$ independent realizations in a system of size $N=10^6$.}
\label{frozen}
\end{figure}

\smallskip
\noindent{\it Frozen clusters.} We now turn to the frozen 
clusters. Their density is found from $dF_k/dt=\alpha c_k$. When
$\alpha\geq \alpha_c$, the density of active clusters over the entire
time range is given by Eq.~(\ref{ck-pre}), and therefore $F_k(t)$ is
found by simple integration. In particular, the final density is
\begin{equation}
\label{Pk-final}
F_k(\infty)=\frac{\alpha}{k^2\cdot
k!}\,\,\gamma(k,k/\alpha)
\end{equation}
where $\gamma(n,x)=\int_0^x dy\,y^{n-1}\,e^{-y}$ is the incomplete
gamma function. At large sizes, the size distribution of frozen
clusters decays according to
\begin{equation}
F_k(\infty)\simeq
\begin{cases}
\frac{1}{2}\cdot k^{-3}                          &\alpha=1\\
A(\alpha)\,k^{-7/2}\exp\left[-B(\alpha) k\right] &\alpha>1
\end{cases}
\end{equation}
where $A=(2\pi)^{-1/2}\alpha^2/(\alpha-1)$ and
$B=\alpha^{-1}+\ln\alpha -1$.

Similarly, we may obtain the density of frozen clusters formed in the
first coagulation case. Integrating (\ref{ck-pre}) gives
\hbox{$F_k(t_g)=\frac{\alpha}{k^2\cdot k!}\,\,\gamma(k,k)$}. This size
distribution decays algebraically, $F_k(t_g)\simeq
\frac{\alpha}{2}k^{-3}$. However, enumerating the frozen clusters that
are born past the first gelation point is more complicated since the
deterministic patches are punctuated by random jumps. In fact, even
the total mass of frozen clusters becomes a random quantity because it
is a complementary quantity to the final mass density of frozen
gels. Interestingly, numerical simulations suggest that the tail
behavior
\begin{equation}
F_k\simeq D\,k^{-3}
\end{equation}
is universal (Fig.~\ref{frozen}). Even the prefactor $D$ appears to
depend only weakly upon the freezing rate $\alpha$.  This shows that
the final distribution of frozen clusters is dominated by the
coagulation phase.  Indeed, in each such phase, the tail behavior is
$k^{-3}$, and furthermore, in the gelation phase, large clusters are
more likely to be consumed by the gel. These results suggest that
freezing leads to an additional non-trivial critical exponent
$\gamma=3$.

In conclusion, we have found that freezing leads to multiple
percolation transitions. The system evolves in a cyclic fashion
alternating between coagulation and gelation. Depending on the
freezing rate, the system may form no gels, a single gel, or multiple
gels. Most statistical characteristics are non-self-averaging because
they are controlled by the random freezing of gels. 

Freezing provides a practical mechanism for controlling gelation. It
may be used to engineer micro-gels of desired size by implementing
variable cooling rates. Slow freezing followed by rapid freezing can
be used to produce gels of prescribed size, while near-critical
freezing produces micro-gels of arbitrarily small size.

There are a number of interesting potential generalizations of the
present work. The most natural is percolation in finite
dimensions. Percolation is analytically tractable only in two
dimensions, though even in that situation it is not clear how to
derive the exponent characterizing the tail of the final distribution
of frozen clusters.  One may also consider situations with different
freezing mechanisms \cite{kb}, particularly for finite and infinite
clusters. We also solved for the case where infinite clusters freeze
immediately \cite{com}. In this case, there is no breakdown of
self-averaging and the final density of frozen clusters mimics the
critical behavior of active clusters, $\gamma=5/2$.

\acknowledgments
We acknowledge US DOE grant W-7405-ENG-36 (EBN) for support of this
work.

\end{document}